\documentclass{article}
\usepackage{upgreek}
\usepackage{latexsym,epsfig,bm,amssymb}
\usepackage{color}
\usepackage{amsthm,mathrsfs}
\usepackage{amsmath}
\usepackage{mathrsfs}
\usepackage{times}

\def\version{March 10, 2008}

\nonstopmode

\newcommand{\notyet}[1]{}

\DeclareSymbolFont{AMSb}{U}{msb}{m}{n}
\DeclareSymbolFontAlphabet{\mathbb}{AMSb}

\newcommand{\lllongrightarrow}{\,-\!\!\!\!-\!\!\!\!-\!\!\!\!-\!\!\!\!\longrightarrow\,}
\newcommand{\E}{\mathscr{E}}
\newcommand{\bS}{\mathbf{S}}

\newcommand\dist{\,{\rm dist}\,}

\newcommand\supp{\mathop{\rm supp}}

\newcommand{\at}[1]{\vert\sb{\sb{#1}}}

\newcommand{\Barr}[1]{\mkern2mu\overline{\mkern-2mu#1\mkern-5mu}\mkern5mu}
\def\Re{{\rm Re\, }}
\def\Im{{\rm Im\,}}
\providecommand\C{\mathbb{C}}
\renewcommand\C{\mathbb{C}}
\newcommand{\R}{\mathbb{R}}
\newcommand{\N}{\mathbb{N}}

\newcommand{\Abs}[1]{\left\vert#1\right\vert}
\newcommand{\abs}[1]{\vert #1 \vert}
\newcommand{\Norm}[1]{\left\Vert #1 \right\Vert}
\newcommand{\norm}[1]{\Vert #1 \Vert}
\newcommand{\const}{{\rm const}\,}

\newcommand\sothat{{\rm :}\ }

\providecommand{\ltor}[1]{
\ifnum #1=1{\it i}\else\ifnum #1=2{\it ii}\else\ifnum #1=3{\it iii}
\else\ifnum #1=4 {\it iv}\fi\fi\fi\fi
}

\DeclareMathSymbol{\varGamma}{\mathord}{letters}{"00}
\DeclareMathSymbol{\varDelta}{\mathord}{letters}{"01}
\DeclareMathSymbol{\varSigma}{\mathord}{letters}{"06}
\DeclareMathSymbol{\varPhi}{\mathord}{letters}{"08}
\DeclareMathSymbol{\varOmega}{\mathord}{letters}{"0A}

\theoremstyle{plain}
\newtheorem{theorem}{Theorem}[section]

\newtheorem{lemma}[theorem]{Lemma}

\newtheorem{proposition}[theorem]{Proposition}
\theoremstyle{definition}
\newtheorem{definition}[theorem]{Definition}
\newtheorem{assumption}{Assumption}

\theoremstyle{remark}
\newtheorem{remark}[theorem]{Remark}

\makeatletter\@addtoreset{equation}{section}
\makeatother

\begin{document}
\title{Global attraction to solitary waves for Klein-Gordon equation
\\
with mean field interaction}

\author{
{\sc Alexander Komech}
\footnote{
Supported in part by Alexander von Humboldt Research Award
(2006),
by DFG grant 436\,RUS\,113/929/0-1, FWF grant P19138-N13,
RFBR grant 07-01-00018a,
and RFBR-DFG grant 08-01-91950-NNIOa.}
\\
{\it\small
Faculty of Mathematics, University of Vienna, Wien A-1090, Austria}
\\
{\it\small
Institute for Information Transmission Problems,
Moscow 101447, Russia}
\\ \\
{\sc Andrew Komech}
\footnote{
Supported in part
by Max-Planck Institute for Mathematics in the Sciences (Leipzig),
Technische Universit\"at M\"unchen,
and by the National Science Foundation under Grant DMS-0600863.
}
\\
{\it\small
Mathematics Department, Texas A\&M University,
College Station, TX, USA}
\\
{\it\small
Institute for Information Transmission Problems,
Moscow 101447, Russia}
}

\date{\version}

\maketitle

\begin{abstract}
We consider a $\mathbf{U}(1)$-invariant nonlinear Klein-Gordon equation
in dimension $n\ge 1$, self-interacting via the mean field mechanism.
We analyze the long-time asymptotics of finite energy solutions
and prove that, under certain generic assumptions,
each solution converges as $t\to\pm\infty$
to the two-dimensional set of all ``nonlinear eigenfunctions''
of the form $\phi(x)e\sp{-i\omega t}$.
This global attraction is caused by the nonlinear energy transfer
from lower harmonics to the continuous spectrum
and subsequent dispersive radiation.
\end{abstract}


\section{Introduction and main results}
\label{sect-results}

In this paper,
we establish the global attraction to the variety of all solitary waves
for the complex Klein-Gordon field $\psi(x,t)$
with the mean field self-interaction:
\begin{equation}\label{kg-mf}
\left\{
\begin{array}{l}
\ddot\psi(x,t)
=\Delta\psi(x,t)-m^2\psi(x,t)
+\rho(x)F(\langle\rho,\psi(\cdot,t)\rangle),
\qquad
x\in\R^n,
\quad n\ge 1,
\quad t\in\R,
\\
\psi\at{t=0}=\psi\sb 0(x),
\qquad
\dot\psi\at{t=0}=\pi\sb 0(x),
\end{array}\right.
\end{equation}
where
\[
\langle\rho,\psi(\cdot,t)\rangle
=\int\sb{\R^n}\bar\rho(x)\psi(x,t)\,d\sp n x.
\]
We assume that $\rho$ is a smooth real-valued
function
from the Schwartz class:
$\rho\in\mathscr{S}(\R^n)$, $\rho\not\equiv 0$.

The long time asymptotics for
nonlinear
wave equations
have been the subject of intensive research,
starting with the pioneering papers by
Segal \cite{MR0153967,MR0152908},
Strauss \cite{MR0233062},
and Morawetz and Strauss \cite{MR0303097},
where the
nonlinear scattering and the local attraction to zero solution
were proved.
Local attraction to solitary waves,
or \emph{asymptotic stability},
in
$\mathbf{U}(1)$-invariant dispersive systems
was addressed in
\cite{MR1071238,MR1199635e,MR1170476,MR1334139}
and then developed in
\cite{MR1488355,MR1681113,MR1893394,MR1835384,MR1972870,MR2027616}.
Global attraction
to \emph{static},
stationary solutions in dispersive systems
\emph{without $\mathbf{U}(1)$ symmetry}
was first established in
\cite{MR1203302e,MR1359949,MR1412428,MR1434147,MR1726676,MR1748357}.

The present paper is our third result
on the global attraction to solitary waves in
$\mathbf{U}(1)$-invariant dispersive systems.
In \cite{ubk-arma}, we proved such an attraction
for the Klein-Gordon field coupled to one nonlinear oscillator.
In \cite{ukk-mpi}, we generalized this result
for the Klein-Gordon field coupled to several oscillators.
Now we are going to extend our theory to a higher dimensional setting,
for the Klein-Gordon equation with the mean field interaction.
This model could be viewed as a generalization
of the $\delta$-function coupling \cite{ubk-arma,ukk-mpi}
to higher dimensions.
We follow the cairns of the approach
we developed in \cite{ubk-arma,ukk-mpi}:
the proof of the absolute continuity of the
spectral density for large frequencies,
the compactness argument to extract the omega-limit trajectories,
and then the usage of the Titchmarsh Convolution Theorem
to pinpoint the spectrum to just one frequency.
The substantial modification is due to
apparent impossibility
to split off a dispersive component and to get the
convergence to the attractor in the local energy norm,
as in \cite{ubk-arma,ukk-mpi};
the convergence which we prove is $\varepsilon$-weaker.
On the other hand, the proof of this slightly weaker convergence
allows us to avoid the technique of quasimeasures,
considerably shortening the argument.

We are aware of only one other recent advance \cite{MR2304091}
in the field
of nonzero global attractors for Hamiltonian PDEs.
In that paper, the global attraction
for the nonlinear Schr\"odinger equation
in dimensions $n\ge 5$ was considered.
The dispersive wave was explicitly specified
using the rapid decay of local energy in higher dimensions.
The global attractor was proved to be compact, but it was
neither identified with the set of solitary waves nor was proved
to be of finite dimension \cite[Remark 1.18]{MR2304091}.

\bigskip

Let us give the plan of the paper.
In the remainder of this section, we formulate
the assumptions and the results.
The proof of the Main Theorem
takes up Section~\ref{sect-bound}
(where we analyze the absolute continuity of the spectrum
for large frequencies)
and Section~\ref{sect-spectral}
(where we select omega-limit trajectories and analyze
their spectrum with the aid of the Titchmarsh Convolution Theorem).
The example of a multifrequency solitary waves
in the situation when $\rho$ is orthogonal to some of the
solitary waves is constructed in Section~\ref{sect-counterexample}.
In Appendix~\ref{sect-existence}  we give a brief sketch
of the proof of the global well-posedness for equation \eqref{kg-mf}.


\subsection{Hamiltonian structure}

We set
$\Psi(t)=
(\psi(x,t),\,\pi(x,t))
$
and rewrite the Cauchy problem \eqref{kg-mf}
in the vector form:
\begin{equation}\label{kg-mf-cp}
\dot\Psi(t)
=
\left[\begin{array}{cc}0&1\\\Delta-m^2&0\end{array}\right]
\Psi(t)
+
\rho(x)\left[\begin{array}{c}0\\F(\langle\rho,\psi(\cdot,t)\rangle)
\end{array}\right],
\qquad
\Psi\at{t=0}
=\Psi\sb 0,
\qquad
x\in\R^n,
\quad n\ge 1,
\quad t\in\R,
\end{equation}
where $\Psi\sb 0=(\psi\sb 0,\pi\sb 0)$.
We assume that
the nonlinearity $F$ admits a real-valued potential:
\begin{equation}\label{P}
F(z)=-\nabla U(z),\quad z\in\C,
\qquad
U\in C\sp 2(\C),
\end{equation}
where the gradient is taken with respect to $\Re z$ and $\Im z$.
Then equation \eqref{kg-mf-cp}
formally can be written as a Hamiltonian system,
\[
\dot\Psi(t)=J\,D\mathcal{H}(\Psi),
\qquad
J=\left[\begin{array}{cc}0&1\\-1&0\end{array}\right],
\]
where $D\mathcal{H}$ is the variational
derivative of the Hamilton functional
\begin{equation}\label{hamiltonian}
\mathcal{H}(\Psi)
=\frac 1 2
\int\sb{\R^n}
\left(
\abs{\pi}\sp 2+\abs{\nabla\psi}\sp 2+m^2\abs{\psi}\sp 2
\right)
d\sp n x
+U(\langle\rho,\psi\rangle),
\quad
\Psi=\left[\begin{array}{c}\psi(x)\\\pi(x)\end{array}\right].
\end{equation}
We
assume that the potential $U(z)$ is $\mathbf{U}(1)$-invariant,
where $\mathbf{U}(1)$ stands for the unitary group
$e\sp{i\theta}$, $\theta\in\R\mod 2\pi$.
Namely, we assume that
there exists $u\in C\sp 2(\R)$ such that
\begin{equation}\label{inv-u}
U(z)=u(\abs{z}\sp 2),
\qquad z\in\C.
\end{equation}
Relations \eqref{P} and \eqref{inv-u}
imply that
\begin{equation}\label{def-a}
F(z)=\alpha(\abs{z}^2)z,
\qquad z\in\C,
\end{equation}
where
$\alpha(\cdot)=-2 u'(\cdot)\in C\sp 1(\R)$
is real-valued.
Therefore,
\begin{equation}\label{inv-f}
F(e\sp{i\theta}z)=e\sp{i\theta} F(z),
\qquad\theta\in\R,\quad z\in\C.
\end{equation}
Due to the $\mathbf{U}(1)$-invariance,
the N\"other theorem formally implies that the functional
\begin{equation}\label{cal-Q}
\mathcal{Q}(\Psi)
=\frac{i}{2}\int\sb{\R^n}
\left(\overline\psi\pi-\overline\pi\psi\right)\,d\sp n x,
\qquad
\Psi=\left[\begin{array}{c}\psi(x)\\\pi(x)\end{array}\right],
\end{equation}
is conserved for solutions $\Psi(t)$ to \eqref{kg-mf-cp}.

We introduce the phase space
of finite energy states for equation \eqref{kg-mf-cp}.
Denote by $\norm{\cdot}\sb{L\sp 2}$
the norm in $L\sp 2(\R^n)$.
Let
$H\sp s(\R^n)$, $s\in\R$,
be the Sobolev space
with the norm
\begin{equation}\label{def-sobolev}
\norm{\psi}\sb{H\sp s}
=\norm{(m^2-\Delta)^{s/2}\psi}\sb{L\sp 2}.
\end{equation}
For $s\in\R$ and $R>0$,
denote by $H\sp s\sb{0}(\mathbb{B}^n\sb R)$
the space of
distributions from $H\sp s(\R^n)$
supported
in $\mathbb{B}^n\sb R$
(the ball of radius $R$ in $\R^n$).
We denote by $\norm{\cdot}\sb{H\sp s,R}$
the norm in the space
$H\sp s(\mathbb{B}^n\sb R)$
which is defined as the dual to
$H\sp{-s}\sb{0}(\mathbb{B}^n\sb R)$.

\begin{definition}
\begin{enumerate}
\item
$\E=H\sp 1(\R^n)\oplus L\sp 2(\R^n)$ is the
Hilbert space
of states
$\Psi
=
\left[\begin{array}{c}\psi(x)
\\
\pi(x)
\end{array}\right]$,
with the norm
\begin{equation}\label{def-e}
\norm{\Psi}\sb{\E}^2
=
\norm{ \pi}\sb{L\sp 2}^2
+\norm{\nabla\psi}\sb{L\sp 2}^2+m^2\norm{\psi}\sb{L\sp 2}^2
=
\norm{ \pi}\sb{L\sp 2}^2
+\norm{\psi}\sb{H\sp 1}^2
.
\end{equation}
\item
For $\varepsilon\ge 0$,
introduce the
Banach spaces
$\E\sp{-\varepsilon}=H\sp{1-\varepsilon}(\R^n)\oplus
H\sp{-\varepsilon}(\R^n)$
with the norm
defined by
\begin{equation}\label{def-e-r-e}
\norm{\Psi }\sb{\E\sp{-\varepsilon}}^2
=
\norm{(m^2-\Delta)^{-\varepsilon/2}\Psi}\sb{\E}^2
=
\norm{\pi}\sb{H\sp{-\varepsilon}}^2
+
\norm{\psi}\sb{H\sp{1-\varepsilon}}^2.
\end{equation}
\item
Define the seminorms
\begin{equation}\label{def-e-r-e-r}
\norm{\Psi }\sb{\E\sp{-\varepsilon},R}^2
=
\norm{\pi}\sb{H\sp{-\varepsilon},R}^2
+
\norm{\psi}\sb{H\sp{1-\varepsilon},R}^2,
\qquad R>0,
\end{equation}
and denote by $\E\sb{loc}\sp{-\varepsilon}$
the space of states $\Psi\in \E\sp{-\varepsilon}$
with {\it finite} norm
\begin{equation}\label{def-e-metric}
\norm{\Psi}\sb{\E\sb{loc}\sp{-\varepsilon}}
=\sum\sb{R=1}\sp{\infty}
2^{-R}\norm{\Psi}\sb{{\E\sp{-\varepsilon},R}}<\infty.
\end{equation}
We will denote $\E\sb{loc}=\E\sb{loc}\sp{0}$.
\end{enumerate}
\end{definition}

\begin{remark}\label{Sob}
The Sobolev embedding theorem implies that the embedding
$\E\subset\E\sb{loc}\sp{-\varepsilon}$
is compact for any $\varepsilon>0$.
\end{remark}

Equation \eqref{kg-mf-cp} is formally the Hamiltonian system
with the phase space $\E$
and the Hamilton functional $\mathcal{H}$.
Both
$\mathcal{H}$
and $\mathcal{Q}$ are continuous functionals on $\E$.
We introduced into \eqref{def-sobolev}
the factor $m^2>0$,
so that
$\mathcal{H}(\Psi)
=\frac 1 2\norm{\Psi}\sb{\E}^2+U(\langle\rho,\psi\rangle)$.

\subsection{Global well-posedness}

To have a priori estimates available for the proof of the global
well-posedness, we assume that
\begin{equation}\label{bound-below}
U(z)\ge {A}-{B}\abs{z}^2
\quad{\rm for}\ z\in\C,\quad
{\rm where}\ {A}\in\R\ {\rm and}
\ 0\le {B}<\frac{m^2}{2\norm{\rho}\sb{L\sp 2}^2}.
\end{equation}

\begin{theorem}\label{theorem-well-posedness}
Let $\rho\in\mathscr{S}(\R^n)$,
and let
$F(z)$ satisfy conditions \eqref{P}, \eqref{inv-u},
and \eqref{bound-below}.
Then:
\begin{enumerate}
\item
For every $\Psi\sb 0\in \E$ the Cauchy problem
\eqref{kg-mf-cp} has a unique solution $\Psi\in C(\R,\E)$.
\item
The map
$W(t):\;\Psi\sb 0\mapsto\Psi(t)$
is continuous in $\E$ and $\E\sb{loc}$
for each $t\in\R$.
\item
The values of the energy and charge functionals are conserved:
\begin{equation}\label{ec}
\mathcal{H}(\Psi(t))=\mathcal{H}(\Psi\sb 0),
\qquad
\mathcal{Q}(\Psi(t))=\mathcal{Q}(\Psi\sb 0),
\qquad
t\in\R.
\end{equation}
\item
The following \emph{a priori} bound holds:
\begin{equation}\label{eb}
\norm{\Psi(t)}\sb{\E}
\le C(\Psi\sb 0)<\infty,
\qquad t\in\R.
\end{equation}
\item
For any $\varepsilon\ge 0$,
the map
$W(t):\Psi\sb 0\mapsto\Psi(t)$
is continuous in $\E\sp{-\varepsilon}$ and $\E\sb{loc}\sp{-\varepsilon}$
uniformly in $t\in[-T,T]$, for any $T>0$.
\end{enumerate}
\end{theorem}

We sketch the proof of this theorem in Appendix~\ref{sect-existence}.

\subsection{Solitary waves}

\begin{definition}
\label{def-solitary-waves}
\begin{enumerate}
\item
The solitary waves of equation \eqref{kg-mf}
are solutions of the form
\begin{equation}\label{solitary-waves}
\psi(x,t)=\phi\sb\omega(x) e\sp{-i\omega t},
\qquad
{\rm where}\
\omega\in\R,
\ \ \phi\sb\omega(x)\in H\sp 1(\R^n).
\end{equation}
\item
The solitary manifold is the set
$
\bS
=
\left\{
(\phi\sb\omega,-i\omega\phi\sb\omega)\sothat\omega\in\R
\right\},
$
where $\phi\sb\omega$ are the amplitudes of solitary waves.
\end{enumerate}
\end{definition}

Identity \eqref{inv-f} implies that the set $\bS$
is invariant under multiplication by $e\sp{i\theta}$,
$\theta\in\R$.
Let us note that since $F(0)=0$ by \eqref{def-a},
for any $\omega\in\R$
there
is a zero solitary wave, $\phi\sb\omega(x)\equiv 0$.

Define
\begin{equation}\label{def-v}
\varSigma(x,\omega)
=\mathcal{F}\sb{\xi\to x}
\Big[\frac{\hat\rho(\xi)}{\xi^2+m^2-\omega^2}\Big],
\qquad
\omega\in\C\sp{+}\cup(-m,m),
\end{equation}
where $\C\sp{+}=\{\omega\in\C\sothat \Im\omega>0\}$.
Note that
$\varSigma(\cdot,\omega)$ is an analytic function of $\omega\in\C\sp{+}$
with the values in $\mathscr{S}(\R^n)$.
Since
$\abs{\varSigma(x,\omega)}\le \const\abs{\Im\omega}^{-1}$
for $\omega\in\C\sp{+}$,
we can extend for any $x\in\R^n$
the function $\varSigma(x,\omega)$ to the entire real line $\omega\in\R$
as a boundary trace:
\begin{equation}\label{def-v-real}
\varSigma(x,\omega)=\lim\sb{\epsilon\to 0+}\varSigma(x,\omega+i\epsilon),
\qquad\omega\in\R,
\end{equation}
where the limit holds in the sense of tempered distributions.

\begin{proposition}[Existence of solitary waves]
\label{prop-solitons}
Assume that $F(z)$ satisfies \eqref{inv-f},
and that $\rho\in\mathscr{S}(\R^n)$, $\rho\not\equiv 0$.
There may only be nonzero solitary wave solutions
to \eqref{kg-mf-cp} for $\omega\in[-m,m]\cup Z\sb\rho$,
where
\begin{equation}\label{def-Omega}
Z\sb\rho
=\{\omega\in\R\backslash[-m,m]\sothat
\hat\rho(\xi)=0\ {\rm for\ all\ }\xi\in\R^n
\ {\rm such\ that\ }m^2+\xi^2=\omega^2\}.
\end{equation}
The profiles of solitary waves
are given by
\[
\hat\phi\sb\omega(\xi)=\frac{c\hat\rho(\xi)}{\xi^2+m^2-\omega^2},
\]
where $c\in\C$, $c\ne 0$ is a root of the equation
\begin{equation}\label{cond-0}
\sigma(\omega)\alpha(\abs{c}^2\abs{\sigma(\omega)}^2)=1,
\end{equation}
where $\alpha$ is defined in \eqref{def-a}
and
\begin{equation}\label{def-sigma}
\sigma(\omega)
=\langle\rho,\varSigma(\cdot,\omega)\rangle
=\frac{1}{(2\pi)^n}
\int\sb{\R^n}\frac{\abs{\hat\rho(\xi)}^2}{\xi^2+m^2-\omega^2}
\,d\sp n\xi.
\end{equation}
The existence of such a root
is a necessary condition
for the existence of nonzero solitary waves \eqref{solitary-waves}.

The condition \eqref{cond-0}
is also sufficient for
$n\ge 5$ and for $\abs{\omega}\ne m$, $n\ge 1$.

For $\abs{\omega}=m$, $n\le 4$,
the following additional condition is needed for sufficiency:
\begin{equation}\label{cond-1}
\int\sb{\R^n}\frac{\abs{\hat\rho(\xi)}^2}{\xi^4}\,d\sp n\xi<\infty.
\end{equation}
\end{proposition}

\begin{remark}\label{remark-sigma}
As follows from \eqref{cond-0} and \eqref{def-sigma},
$\sigma(\omega)$ is strictly positive for $\abs{\omega}<m$
(since $\rho\not\equiv 0$)
and takes finite nonzero values for all $\omega$
that correspond to solitary waves
(for $n\le 4$, the finiteness of $\sigma(\omega)$
at $\omega=\pm m$
follows if \eqref{cond-1} is satisfied).
\end{remark}

\begin{remark}
One can see that generically the solitary wave manifold is two-dimensional.
\end{remark}

\begin{proof}
Substituting the ansatz $\phi\sb\omega(x) e^{-i\omega t}$
into \eqref{kg-mf} and using \eqref{def-a},
we get the following equation on $\phi\sb\omega$:
\begin{equation}\label{stat-eqn}
-\omega^2\phi\sb\omega(x)
=
\Delta\phi\sb\omega(x)
-m^2\phi\sb\omega(x)
+\rho(x)F(\langle\rho,\phi\sb\omega\rangle),
\qquad x\in\R^n.
\end{equation}
Therefore, all solitary waves satisfy the relation
\begin{equation}\label{phi-phi}
(\xi^2+m^2-\omega^2)\hat\phi\sb\omega(\xi)
=\hat\rho(\xi)F(\langle\rho,\phi\sb\omega\rangle).
\end{equation}
For $\omega\notin[-m,m]\cup Z\sb\rho$
the relation \eqref{phi-phi}
leads to $\phi\sb\omega\notin L\sp 2(\R^n)$
(unless $\phi\sb\omega\equiv 0$).
We conclude that there are no nonzero solitary waves for
$\omega\notin[-m,m]\cup Z\sb\rho$.

Let us consider the case $\omega\in[-m,m]\cup Z\sb\rho$.
From \eqref{phi-phi}, we see that
\begin{equation}\label{phi-phi-2}
\hat\phi\sb\omega(\xi)
=\frac{\hat\rho(\xi)}{\xi^2+m^2-\omega^2}
F(\langle\rho,\phi\sb\omega\rangle).
\end{equation}
Using the function $\varSigma(x,\omega)$ defined in \eqref{def-v},
we may express
$\phi\sb\omega(x)=c \varSigma(x,\omega)$,
with $c\in\C$.
Substituting
this ansatz into \eqref{phi-phi-2}
and using \eqref{def-a}, we can write the condition on $c$
in the form \eqref{cond-0}.

For $n\le 4$, the finiteness of the energy of solitons
corresponding to $\omega=\pm m$
is equivalent to the condition \eqref{cond-1}.

This finishes the proof of the proposition.
\end{proof}

\subsection{The main result}

\begin{assumption}\label{ass-rho}
We assume that
$\rho\in\mathscr{S}(\R^n)$,
the set
$Z\sb\rho$
is finite,
and that
\begin{equation}
\sigma(\omega)
\ne 0,
\qquad
\omega\in Z\sb\rho.
\end{equation}
Above,
$Z\sb\rho$ and $\sigma(\omega)$
are defined in \eqref{def-Omega} and \eqref{def-sigma}.
\end{assumption}

\begin{remark}
Note that $\sigma(\omega)$ is well-defined
at the points of $Z\sb\rho$
since
$\hat\rho\big\vert\sb{\abs{\xi}=\sqrt{\omega^2-m^2}}\equiv 0$
for $\omega\in Z\sb\rho$.
\end{remark}

As we mentioned before,
we need to assume that the nonlinearity is polynomial.
This assumption
is crucial in our argument:
It will allow us to apply the Titchmarsh Convolution Theorem.
Now all our assumptions on $F$
can be summarised as follows.

\begin{assumption}\label{ass-f}
$F(z)$ satisfies \eqref{P} with the polynomial potential $U(z)$,
and also satisfies \eqref{inv-u} and \eqref{bound-below}.
This can be summarised as the following assumption on $U(z)$:
\begin{equation}\label{f-is-such}
U(z)=\sum\limits\sb{n=1}\sp{p}u\sb n\abs{z}\sp{2n},
\qquad
u\sb n\in\R,
\quad
p\ge 2,
\quad
u\sb p>0.
\end{equation}
\end{assumption}

Our main result is the following theorem.

\begin{theorem}[Main Theorem]
\label{main-theorem}
Assume that
the coupling function $\rho(x)$ satisfies Assumption~\ref{ass-rho}
and that
the nonlinearity $F(z)$ satisfies Assumption~\ref{ass-f}.
Then for any $\Psi\sb 0\in \E$
the solution $\Psi(t)\in C(\R,\E)$
to the Cauchy problem \eqref{kg-mf-cp}
converges to $\bS$ in the space
$\E\sb{loc}\sp{-\varepsilon}$,
for any $\varepsilon>0$:
\begin{equation}\label{cal-A}
\lim\sb{t\to\pm\infty}
\dist\sb{\E\sb{loc}\sp{-\varepsilon}}(\Psi(t),\bS)=0,
\end{equation}
where
$\dist\sb{\E\sb{loc}\sp{-\varepsilon}}(\Psi,\bS)
:=\inf\limits\sb{\bm s\in\bS}
\norm{\Psi-\bm s}\sb{\E\sb{loc}\sp{-\varepsilon}}$.
\end{theorem}

\begin{remark}
The $\E\sb{loc}\sp{-\varepsilon}$-convergence
to the attractor stated in this theorem
is weaker than the $\E\sb{loc}$-convergence
proved in \cite{ubk-arma} and \cite{ukk-mpi},
where we considered the Klein-Gordon field
in dimension $n=1$,
coupled to nonlinear oscillators.
\end{remark}

Obviously,
it suffices to prove Theorem~\ref{main-theorem}
for $t\to+\infty$.

\section{Absolute continuity for large frequencies}
\label{sect-bound}

\subsection{Splitting of a dispersive component}

First we split the solution
$\psi(x,t)$
into
$\psi(x,t)=\chi(x,t)+\varphi(x,t)$,
where
$\chi$
and
$\varphi$
are defined
as solutions to the following Cauchy problems:
\begin{eqnarray}
&&
\ddot\chi(x,t)
=
\Delta\chi(x,t)-m^2\chi(x,t),
\qquad
(\chi,\dot\chi)\at{t=0}=\Psi\sb 0,
\label{kg-mf-cp-1}
\\
\nonumber
\\
&&
\ddot\varphi(x,t)
=
\Delta\varphi(x,t)-m^2\varphi(x,t)
+\rho(x)f(t),
\qquad
(\varphi,\dot\varphi)\at{t=0}=(0,0),
\label{kg-mf-cp-2}
\end{eqnarray}
where
$\Psi\sb 0$
is the initial data from
\eqref{kg-mf-cp},
and
\begin{equation}\label{def-f}
f(t):=F(\langle\rho,\psi(\cdot,t)\rangle).
\end{equation}
Note that
$\langle\rho,\psi(\cdot,t)\rangle$
belongs to $C\sb{b}(\R)$
since
$(\psi,\dot\psi)\in C\sb{b}(\R, \E)$
by Theorem~\ref{theorem-well-posedness}~({\it iv}).
Hence,
\begin{equation}\label{f-is-cb}
f(\cdot)\in C\sb{b}(\R).
\end{equation}
On the other hand,
since $\chi(t)$
is a finite energy solution to the free Klein-Gordon equation,
we also have
\begin{equation}\label{psi-1-bounds}
(\chi,\dot\chi)
\in C\sb{b}(\R, \E).
\end{equation}
Hence, the function $\varphi(t)=\psi(t)-\chi(t)$
also satisfies
\begin{equation}\label{psi-2-bounds}
(\varphi,\dot\varphi)
\in C\sb{b}(\R,\E).
\end{equation}

The following lemma reflects the well-known energy decay
for the linear Klein-Gordon equation.

\begin{lemma}\label{lemma-decay-psi1}
There is a local decay of $\chi$
in the $\E\sb{loc}$ seminorms.
That is,
$\forall R>0$,
\begin{equation}\label{dp0}
\Norm{(\chi(t),\dot\chi(t))}\sb{\E,R}
\to 0,
\qquad t\to\infty.
\end{equation}
\end{lemma}

\subsection{Complex Fourier-Laplace transform}

Let us analyze the complex Fourier-Laplace transform of
$\varphi(x,t)$:
\begin{equation}\label{FL}
\displaystyle
\tilde\varphi(x,\omega)
=\mathcal{F}\sb{t\to\omega}[\Theta(t)\varphi(x,t)]
:=
\int\sb 0\sp\infty e\sp{i\omega t}\varphi(x,t)\,dt,
\quad\omega\in\C\sp{+},
\quad
x\in\R\sp n,
\end{equation}
where
$\C\sp{+}:=\{z\in\C:\;\Im z>0\}$.
Due to \eqref{psi-2-bounds},
$\tilde\varphi(\cdot,\omega)$
is an $H\sp 1$-valued analytic function of $\omega\in\C\sp{+}$.
Equation \eqref{kg-mf-cp-2}
for $\varphi$ implies that
\[
-\omega^2\tilde\varphi(x,\omega)
=
\Delta\tilde\varphi(x,\omega)-m^2\tilde\varphi(x,\omega)
+\rho(x)\tilde f(\omega),\quad \omega\in\C\sp{+},
\quad x\in\R\sp n,
\]
where $\tilde f(\omega)$ is the Fourier-Laplace
transform of $f(t)$:
\[
\tilde f(\omega)
=\mathcal{F}\sb{t\to\omega}[\Theta(t)f(t)]
=\int\sb{0}\sp\infty e^{i\omega t}f(t)\,dt,
\qquad\omega\in\C\sp{+}.
\]
The solution $\tilde\varphi(x,\omega)$
is analytic for $\omega\in\C\sp{+}$
and can be represented by
\begin{equation}\label{psi-s}
\tilde\varphi(x,\omega)
=\varSigma(x,\omega)\tilde f(\omega),
\qquad\omega\in\C\sp{+}.
\end{equation}

\subsection{Traces of distributions for $\omega\in\R$}

First we remark that
\begin{equation}\label{psi2}
\Theta(t)\varphi(x,t)\in C\sb{b}(\R, H\sp 1(\R^n))
\end{equation}
by \eqref{psi-2-bounds} since $\varphi(x,0+)=0$ by
initial conditions in \eqref{kg-mf-cp-2}.
The Fourier-Laplace transform of $\varphi$ in time,
$
\mathcal{F}\sb{t\to\omega}[\Theta(t)\varphi(\cdot,t)]$,
is a tempered $H\sp 1$-valued distribution of $\omega\in\R$
by \eqref{psi-2-bounds}.
We will denote this Fourier-Laplace transform by
$\tilde\varphi(\cdot,\omega)$,
$\omega\in\R$,
which is the boundary value of
the analytic function $\tilde \varphi(\cdot,\omega)$, $\omega\in\C\sp{+}$,
in the following sense:
\begin{equation}\label{bvp1}
\tilde\varphi(\cdot,\omega)
=\lim\limits\sb{\epsilon\to 0+}\tilde\varphi(\cdot,\omega+i\epsilon),
\qquad\omega\in\R,
\end{equation}
where the convergence is in the space of
$H\sp 1$-valued tempered distributions of $\omega$,
$\mathscr{S}'(\R,H\sp 1(\R^n))$.
Indeed,
\[
\tilde\varphi(\cdot,\omega+i\epsilon)
=\mathcal{F}\sb{t\to\omega}[\Theta(t)\varphi(\cdot,t)e\sp{-\epsilon t}],
\]
while
$\Theta(t)\varphi(\cdot,t)e\sp{-\epsilon t}
\mathop{\longrightarrow}\limits\sb{\epsilon\to 0+}
\Theta(t)\varphi(\cdot,t),
$
with the convergence taking place in
$\mathscr{S}'(\R,H\sp 1(\R^n))$
which is the space of $H\sp 1$-valued tempered distributions of $t\in\R$.
Therefore, \eqref{bvp1} holds by the continuity of the Fourier transform
$\mathcal{F}\sb{t\to\omega}$ in $\mathscr{S}'(\R)$.
Similarly to \eqref{bvp1},
the distribution $\tilde f(\omega)$ for $\omega\in\R$
is the boundary value of the analytic in $\C\sp{+}$
function $\tilde f(\omega)$, $\omega\in\C\sp{+}$:
\begin{equation}\label{bv}
\tilde f(\omega)=\lim\limits\sb{\epsilon\to 0+}
\tilde f(\omega+i\epsilon), \quad \omega\in\R,
\end{equation}
since the function
$\Theta(t)f(t)$ is bounded.
The convergence holds in the space of tempered distributions
$\mathscr{S}'(\R)$.

Let us justify that the representation \eqref{psi-s}
for $\tilde\varphi(x,\omega)$
is also valid when $\omega\in\R$, $\omega\ne\pm m$,
if the multiplication in \eqref{psi-s}
is understood in the sense of distributions.

\begin{proposition}\label{prop-uniform}
For any fixed $x\in\R^n$,
$\varSigma(x,\omega)$, $\omega\in\R\backslash\{\pm m\}$,
is a smooth function,
and the identity
\begin{equation}\label{p1r}
\tilde\varphi(x,\omega)=\varSigma(x,\omega)\tilde f(\omega),
\qquad\omega\in\R\backslash\{\pm m\},
\end{equation}
holds in the sense of distributions.
\end{proposition}

\begin{proof}
Consider
\begin{equation}
\varSigma(x,\omega)
=\frac{1}{(2\pi)^n}\int\sb{\R^n}
\frac{e^{i\xi x}\hat\rho(\xi)\,d\sp n\xi}{\xi^2+m^2-(\omega+i0)^2}
=\int\sb{0}\sp{\infty}
\frac{R(x,\eta)\,d\eta}{\eta^2+m^2-(\omega+i0)^2},
\end{equation}
where
\begin{equation}
R(x,\eta)
=\frac{1}{(2\pi)^n}\int\sb{\abs{\xi}=\eta}
e^{i\xi x}\hat\rho(\xi)
\,d\sp{n-1}S\sb{\xi}.
\end{equation}
For each $x\in\R^n$,
$R(x,\eta)$ is smooth for $\eta>0$
and satisfies
$\abs{R(x,\eta)}=O(\eta^{n-1})$.
It follows that for each $x\in\R^n$,
$\varSigma(x,\omega)$ is a smooth function
of $\omega\in\R\backslash\{\pm m\}$,
and hence is a multiplicator in the space of distributions.
\end{proof}

\subsection{Absolutely continuous spectrum}

Let
$k(\omega)$
denote the branch of
$\sqrt{\omega^2-m^2}$
such that $\Im\sqrt{\omega^2-m^2}\ge 0$
for $\omega\in\C\sp{+}$:
\begin{equation}\label{def-k}
k(\omega)=\sqrt{\omega^2-m^2},
\qquad\Im k(\omega)>0,
\qquad
\omega\in\C\sp{+}.
\end{equation}
Then $k(\omega)$ is the analytic function
for $\omega\in\C\sp{+}$.
We extend it to $\omega\in\Barr{\C\sp{+}}$ by continuity.

\begin{proposition}\label{tilde-f-plus-bounded}
The distribution $\tilde f(\omega+i0)$, $\omega\in\R$,
is absolutely continuous for $\abs{\omega}>m$
and satisfies
\begin{equation}\label{rho-f-bound-we-have}
\int\sb{\abs{\omega}>m}
\abs{\tilde f(\omega)}^2
\mathscr{M}(\omega)
\,d\omega
<\infty,
\end{equation}
where
$\mathscr{M}(\omega)
=\frac{1}{\omega^2}\mathscr{R}(\abs{k(\omega)})$,
$\ \mathscr{R}(\eta)
=\frac{1}{(2\pi)^n}\int\sb{\abs{\xi}=\eta}
\abs{\hat\rho(\xi)}^2
\,d\sp{n-1}S\sb{\xi}$,
$\ \eta>0$.
\end{proposition}

\begin{remark}
The function $\mathscr{M}(\omega)$,
$\abs{\omega}>m$,
is non-negative,
and
its set of zeros
coincides with $Z\sb\rho$ defined in \eqref{def-Omega}.
\end{remark}

\begin{remark}
Recall that
$\tilde f(\omega)$,
$\omega\in\R$,
is defined by \eqref{bv}
as the trace
distribution:
$\tilde f(\omega)=\tilde f(\omega+i0)$.
\end{remark}

\begin{proof}
We will prove that for any
compact interval $I$ such that $I\cap([-m,m]\cup Z\sb\rho)=\emptyset$
the following inequality holds:
\begin{equation}\label{rho-f-bound-we-have-1}
\int\limits\sb{I}
\abs{\tilde f(\omega)}^2
\mathscr{M}(\omega)
\,d\omega
\le C,
\end{equation}
for some constant $C>0$ which does not depend on $I$.
Since there is a finite number of connected components
of $\R\backslash([-m,m]\cup Z\sb\rho)$,
this will finish the proof of the proposition.
Let us prove \eqref{rho-f-bound-we-have-1}.
The Parseval identity applied to
\[
\tilde\varphi(x,\omega+i\epsilon)
=\int\sb{0}\sp\infty
\varphi(x,t)e^{i\omega t-\epsilon t}
\,dt,
\qquad
\omega\in\R,
\quad
\epsilon>0,
\]
leads to
\[
\int\sb{\R}
\norm{\tilde\varphi(\cdot,\omega+i\epsilon)}\sb{L\sp 2}^2\,d\omega
=
2\pi
\int\sb{0}\sp\infty
\norm{\varphi(\cdot,t)}\sb{L\sp 2}^2 e^{-2\epsilon t}\,dt.
\]
Since $\sup\sb{t\ge 0}\norm{\varphi(\cdot,t)}\sb{L\sp 2}<\infty$
by \eqref{psi-2-bounds}, we may bound the right-hand side
by $C\sb 1/\epsilon$, with some $C\sb 1>0$.
Taking into account \eqref{psi-s},
we arrive at the key inequality
\begin{equation}\label{e-plus-bounds}
\int\sb{\R}
\abs{\tilde f(\omega+i\epsilon)}^2
\norm{\varSigma(\cdot,\omega+i\epsilon)}\sb{L\sp 2}^2
\,d\omega
\le
\frac{C\sb 1}{\epsilon}.
\end{equation}

\begin{lemma}\label{lemma-h1-bounds}
Assume that $I$ is a compact interval
such that $I\cap([-m,m]\cup Z\sb\rho)=\emptyset$.
Then there exists $\epsilon\sb I>0$ such that
\begin{equation}\label{n-half}
\norm{\varSigma(\cdot,\omega+i\epsilon)}\sb{L\sp 2}^2
\ge
\frac{\mathscr{M}(\omega)}{40\epsilon},
\qquad\omega\in I,
\quad
0<\epsilon\le\epsilon\sb I.
\end{equation}
\end{lemma}

\begin{proof}
Let us compute the $L\sp 2$-norm
using the Fourier space representation.
Since
$
\hat\varSigma(\xi,\omega+i\epsilon)
=\frac{\hat\rho(\xi)}
{\xi^2+m^2-(\omega+i\epsilon)^2},
$
we have:
\begin{equation}\label{n-v-e}
\norm{\varSigma(\cdot,\omega+i\epsilon)}\sb{L\sp 2}^2
=
\frac{1}{(2\pi)^n}
\int\limits\sb{\R^n}
\frac{\abs{\hat\rho(\xi)^2}\,d\sp n\xi}
{\abs{\xi^2+m^2-(\omega+i\epsilon)^2}^2}
=\int\limits\sb{0}\sp\infty
\frac{\mathscr{R}(\eta)\,d\eta}
{\abs{\eta^2+m^2-(\omega+i\epsilon)^2}^2}.
\end{equation}
Let $K\sb{I}$ be given by
\begin{equation}
K\sb{I}=\{\eta>0\sothat\eta^2=\omega^2-m^2,\ \omega\in I\}.
\end{equation}
We denote
\begin{equation}\label{def-eta-omega}
\eta\sb\omega=\abs{k(\omega)}\in K\sb{I}.
\end{equation}
Since the function $\mathscr{R}(\eta)$
is smooth and strictly positive on $K\sb{I}$,
there exists $\epsilon\sb I>0$,
satisfying
\begin{equation}\label{epsilon-i-small}
\epsilon\sb I<\min(m,\frac 1 2\abs{K\sb{I}}),
\end{equation}
so that
$
\mathscr{R}(\eta\sb 2)
\ge \frac 1 2 \mathscr{R}(\eta\sb 1),
$
for all $\eta\sb 1,\,\eta\sb 2\in K\sb{I}$
such that $\abs{\eta\sb 2-\eta\sb 1}<\epsilon\sb I$.
Hence, \eqref{n-v-e} yields
\begin{equation}\label{n-v-e-2}
\norm{\varSigma(\cdot,\omega+i\epsilon)}\sb{L\sp 2}^2
\ge
\frac{\mathscr{R}(\eta\sb\omega)}{2}
\int\limits\sb{K\sb{I}
  \cap[\eta\sb\omega-\epsilon,\eta\sb\omega+\epsilon]}
\frac{d\eta}{\abs{\eta^2+m^2-(\omega+i\epsilon)^2}^2},
\qquad
0<\epsilon\le\epsilon\sb I.
\end{equation}
Estimating the integral in the right-hand side of
\eqref{n-v-e-2}
via the inequality
\[
\inf\sb{\abs{\eta-\eta\sb\omega}\le\epsilon}
\frac{1}{{\abs{\eta^2+m^2-(\omega+i\epsilon)^2}^2}}
=
\frac{1}{(2\eta\sb\omega\epsilon+2\epsilon^2)^2+4\omega^2\epsilon^2}
\ge
\frac{1}{20\epsilon^2\omega^2},
\]
where we took into account that
$\eta\sb\omega^2=\omega^2-m^2$
and
$\epsilon\le\epsilon\sb I<m<\abs{\omega}$,
we arrive at
\begin{equation}\label{n-v-e-3}
\norm{\varSigma(\cdot,\omega+i\epsilon)}\sb{L\sp 2}^2
\ge
\frac{\mathscr{R}(\eta\sb\omega)}{40\epsilon^2\omega^2}
\Abs{K\sb{I}\cap[\eta\sb\omega-\epsilon,\eta\sb\omega+\epsilon]}
\ge
\frac{\mathscr{R}(\eta\sb\omega)}{40\epsilon\omega^2},
\qquad \omega\in I,
\quad 0<\epsilon\le\epsilon\sb I.
\end{equation}
The last inequality
follows since by
\eqref{def-eta-omega}
and
\eqref{epsilon-i-small}
either
$[\eta\sb\omega-\epsilon,\eta\sb\omega]\subset K\sb{I}$
or $[\eta\sb\omega,\eta\sb\omega+\epsilon]\subset K\sb{I}$
or both.
\end{proof}

Substituting \eqref{n-half} into \eqref{e-plus-bounds},
we obtain the bound
\begin{equation}\label{fin}
\int\sb{I}
\abs{\tilde f(\omega+i\epsilon)}^2
\mathscr{M}(\omega)
\,d\omega
\le 40 C\sb 1,
\qquad
0<\epsilon\le\epsilon\sb I.
\end{equation}
We conclude that the set of functions
$
g\sb{I,\epsilon}(\omega)
=
\tilde f(\omega+i\epsilon)
\sqrt{\mathscr{M}(\omega)},
$
$
0<\epsilon\le\epsilon\sb I,
$
defined for $\omega\in I$,
is bounded in the Hilbert space $L\sp 2(I)$,
and, by the Banach Theorem, is weakly compact.
The convergence of the distributions \eqref{bv}
implies the weak convergence
$g\sb{I,\epsilon}
\mathop{-\!\!\!\!-\!\!\!\!\rightharpoondown}\limits\sb{\epsilon\to 0+}
g\sb{I}$
in the Hilbert space
$L\sp 2(I)$.
The limit function
$g\sb{I}(\omega)$ coincides with the distribution
$\tilde f(\omega)\sqrt{\mathscr{M}(\omega)}$
restricted onto
$I$.
This proves the bound \eqref{rho-f-bound-we-have-1}
and finishes the proof of the proposition.
\end{proof}

\section{Nonlinear spectral analysis of omega-limit trajectories}
\label{sect-spectral}

\subsection{Compactness argument and omega-limit trajectories}

Fix $\Psi\sb 0=(\psi\sb 0,\pi\sb 0)\in\E$,
and let $\psi\in C(\R,H\sp{1}(\R^n))$
be the solution to the Cauchy problem
\eqref{kg-mf} with the initial data
$(\psi,\dot\psi)\at{t=0}=\Psi\sb 0$.
Let $t\sb j>0$, $j\in\N$ be a sequence such that
$t\sb j\to\infty$.

Since $(\psi,\dot\psi)\at{t\sb j}$ are bounded in $\E$,
we can pick a subsequence of $\{t\sb j\}$, also denoted $\{t\sb j\}$,
such that
\begin{equation}\label{psi-tj-to-b}
(\psi,\dot\psi)\at{t\sb j}\mathop{\lllongrightarrow}\sb{j\to\infty} B\sb 0
\qquad
{\rm in}
\ \ \E\sb{loc}\sp{-\varepsilon},
\ \ {\rm for\ any}
\ \ \varepsilon>0,
\end{equation}
where $B\sb 0$ is some vector from $\E$ (see Remark~\ref{Sob}).
By Theorem~\ref{theorem-well-posedness},
there is a solution
$\beta(x,t)\in C(\R,H\sp 1(\R^n))$
to \eqref{kg-mf}
with the initial data $(\beta,\dot\beta)\at{t=0}=B\sb 0\in\E$:
\begin{equation}\label{kg-mf-beta}
\ddot\beta(x,t)
=\Delta\beta(x,t)-m^2\beta(x,t)+\rho(x)F(\langle\rho,\beta\rangle),
\qquad x\in\R^n,\ t\in\R;
\qquad
(\beta,\dot\beta)\at{t=0}=B\sb 0\in\E;
\end{equation}
this solution satisfies the bound
\begin{equation}\label{beta-beta}
\sup\limits\sb{t\in\R}
\norm{(\beta(\cdot,t),\dot\beta(\cdot,t))}\sb{\E}<\infty.
\end{equation}
Let $S\sb{\tau}$ be the time shift operators,
$S\sb{\tau}f(t)=f(t+\tau)$.
By
\eqref{psi-tj-to-b}
and Theorem~\ref{theorem-well-posedness}~({\it v}),
for any $T>0$ and $\varepsilon>0$,
there is the convergence
\begin{equation}\label{olpd0}
S\sb{t\sb j}(\psi,\dot\psi)
\mathop{\lllongrightarrow}\sb{j\to\infty}
(\beta,\dot\beta)
\qquad
{\rm in}
\ \ C\sb{b}([-T,T],\E\sb{loc}\sp{-\varepsilon}).
\end{equation}
If a function $\beta(x,t)$
appears as the limit in \eqref{olpd0}
for some sequence $t\sb j\to\infty$,
we will call it
{\it omega-limit trajectory}.

To conclude the proof of Theorem~\ref{main-theorem},
it suffices to check that every omega-limit trajectory
belongs to the set of solitary waves;
that is,
\begin{equation}\label{claim-beta}
\beta(x,t)
=
\phi\sb{\omega\sb{+}}(x)e\sp{-i\omega\sb{+}t},
\qquad
x\in\R^n,
\quad
t\in\R,
\end{equation}
with some $\omega\sb{+}\in\R$.

\subsection{Compactness of the spectrum}

We denote
$g(t)=F(\langle\rho,\beta(\cdot,t)\rangle)$.

\begin{proposition}\label{prop-beta}
$\supp\tilde g\subset [-m,m]\cup Z\sb\rho$,
where $Z\sb\rho$ is defined in \eqref{def-Omega}.
\end{proposition}

\begin{proof}
By Lemma~\ref{lemma-decay-psi1},
\begin{equation}\label{chi-decay}
\norm{(\chi,\dot\chi)\at{t}}\sb{\E\sb{loc}}
\mathop{\lllongrightarrow}\sb{t\to\infty} 0,
\end{equation}
hence the long-time asymptotics
of the solution $\psi(x,t)$ in $\E\sb{loc}$
depends only on the singular component
$\varphi(x,t)$.
The convergence \eqref{olpd0},
together with \eqref{chi-decay},
prove that
for any $T>0$ and $\varepsilon>0$,
\begin{equation}\label{olpd1}
S\sb{t\sb j}(\varphi,\dot\varphi)
\mathop{\lllongrightarrow}\sb{j\to\infty}(\beta,\dot\beta),
\qquad
{\rm in}
\ \ C\sb{b}([-T,T],\E\sb{loc}\sp{-\varepsilon}).
\end{equation}
The convergence \eqref{olpd1} implies that,
for any smooth compactly supported function
$\alpha(x)$,
there is a convergence
\[
\langle\alpha,\varphi(\cdot,t+t\sb j)\rangle
\mathop{\stackrel{\mathscr{S}'}{\lllongrightarrow}}\sb{j\to\infty}
\langle\alpha,\beta(\cdot,t)\rangle.
\]
Due to the continuity of the Fourier transform from $\mathscr{S}'(\R)$
into itself, we also have
\begin{equation}\label{dztA-alpha}
\zeta(\omega)
\langle\alpha,\tilde\varphi(\cdot,\omega)\rangle
e^{-i\omega t\sb j}
\mathop{\stackrel{\mathscr{S}'}{\lllongrightarrow}}\sb{j\to\infty}
\zeta(\omega)\langle\alpha,\tilde\beta(\cdot,\omega)\rangle,
\end{equation}
where $\zeta(\omega)$ is a smooth compactly supported function.
Assume that $\supp\zeta\cap([-m,m]\cup Z\sb\rho)=\emptyset$.
Then, by Proposition~\ref{prop-uniform},
we may substitute $\zeta(\omega)\tilde\varphi(x,\omega)$
by $\zeta(\omega)\varSigma(x,\omega)\tilde f(\omega)$,
getting
\begin{equation}\label{dztAd}
\zeta(\omega)
\langle\alpha,\varSigma(\cdot,\omega)\rangle
\tilde f(\omega)
e^{-i\omega t\sb j}
\mathop{\stackrel{\mathscr{S}'}\lllongrightarrow}\sb{j\to\infty}
\zeta(\omega)\langle\alpha,\tilde\beta(\cdot,\omega)\rangle.
\end{equation}
Since $\tilde f$ is locally $L\sp 2$ on $\R\backslash([-m,m]\cup Z\sb\rho)$
by Proposition~\ref{tilde-f-plus-bounded},
while  $\varSigma(x,\omega)$ is smooth in $\omega\in\R\backslash\{\pm m\}$
for any $x\in\R^n$,
the product
$\zeta(\omega)\langle\alpha,\varSigma(\cdot,\omega)\rangle\tilde f(\omega)$
is in $L\sp 1(\R)$.
Therefore the left-hand side of \eqref{dztAd} converges to zero.
It follows that
$\tilde\beta(x,\omega)\equiv 0$
for $\omega\notin [-m,m]\cup Z\sb\rho$.
\end{proof}

\subsection{Spectral inclusion}

\begin{proposition}
\label{prop-b-g}
$\supp\tilde g\subset\supp\langle\rho,\tilde\beta(\cdot,\omega)\rangle$.
\end{proposition}

This proposition states that the time spectrum of
$g(t)=F(\langle\rho,\beta(\cdot,t)\rangle)$
is included in the time spectrum of $\langle\rho,\beta(\cdot,t)\rangle$.
This spectral inclusion plays the key role in the proof of our main
result (Theorem~\ref{main-theorem}).

\begin{proof}
By \eqref{olpd1},
\[
f(t+t\sb j)=F(\langle\rho,\varphi(\cdot,t+t\sb j)\rangle)
\mathop{\stackrel{C\sb{b}([-T,T])}{\lllongrightarrow}}\sb{j\to\infty}
F(\langle\rho,\beta(\cdot,t)\rangle)=g(t),
\]
for any $T>0$.
Using \eqref{p1r} and taking into account
that $\varSigma(x,\omega)$ is smooth for $\omega\not=\pm m$,
we obtain the following relation
which holds in the sense of distributions:
\begin{equation}\label{ber}
\tilde\beta(x,\omega)=\varSigma(x,\omega)\tilde g(\omega),
\qquad x\in\R^n,
\qquad\omega\in\R\backslash\{\pm m\}.
\end{equation}
Taking the pairing of \eqref{ber} with $\rho$
and using definition of $\sigma(\omega)$ (see \eqref{def-sigma}), we get:
\begin{equation}\label{ber-rho}
\langle\rho,\tilde\beta(\cdot,\omega)\rangle
=\sigma(\omega)\tilde g(\omega),
\qquad
\omega\in\R\backslash\{\pm m\}.
\end{equation}

First we prove Proposition~\ref{prop-b-g}
modulo the set $\omega=\{\pm m\}$.

\begin{lemma}\label{lemma-same-supp}
$\supp\tilde g\backslash\{\pm m\}
\subset
\supp\langle\rho,\tilde\beta(\cdot,\omega)\rangle$.
\end{lemma}

\begin{proof}
By Proposition~\ref{prop-beta},
$\supp\tilde g\subset [-m,m]\cup Z\sb\rho$.
Thus, the statement of the lemma follows from \eqref{ber-rho}
and from noticing that
$\sigma(\omega)$
is smooth and positive for $\omega\in(-m,m)$
and moreover, by Assumption~\ref{ass-rho},
it is nonzero on $Z\sb\rho$.
\end{proof}

To finish the proof of Proposition~\ref{prop-b-g},
it remains to consider the contribution of $\omega=\pm m$.

\begin{lemma}\label{lemma-m}
If $\omega\sb 0=\pm m$ belongs to $\supp\tilde g$,
then $\omega\sb 0\in\supp\langle\rho,\tilde\beta\rangle$.
\end{lemma}

\begin{proof}
In the case when $\omega\sb 0=\pm m$
is not an isolated point in $[-m,m]\cap\supp\tilde g$,
we use
\eqref{ber-rho}
to conclude that $\omega\sb 0\in \supp\langle\rho,\tilde\beta\rangle$
due to positivity of $\sigma(\omega)$ for
$\abs{\omega}<m$ (which is apparent from \eqref{def-sigma}).

We are left to consider
the case when $\omega\sb 0=m$ or $-m$
is an isolated point in $[-m,m]\cap\supp\tilde g$.
We can pick an open neighbourhood $U$ of $\omega\sb 0$
such that $U\cap\supp\tilde g=\{\omega\sb 0\}$
since
$\supp\tilde g\in[-m,m]\cup Z\sb\rho$
and $Z\sb\rho$ is a discrete finite set.
Pick $\zeta\in C\sp\infty\sb{0}(\R)$, $\supp\zeta\subset U$,
such that $\zeta(\omega\sb 0)=1$.
First we note that
\begin{equation}\label{g-at-u}
\zeta(\omega)\tilde g(\omega)=M\,\delta(\omega-\omega\sb 0),
\qquad
M\in\C\backslash\{0\},
\end{equation}
where the derivatives of the $\delta(\omega-\omega\sb 0)$
are prohibited because $\check\zeta\ast g(t)$ is bounded.
By \eqref{ber},
we have $U\cap \supp\sb{\omega}\tilde\beta\subset\{\omega\sb 0\}$,
hence
\begin{equation}\label{b-b}
\zeta(\omega)\tilde\beta(x,\omega)=\delta(\omega-\omega\sb 0)b(x),
\qquad b\in H\sp 1(\R^n).
\end{equation}
Again,
the terms with the derivatives of $\delta(\omega-\omega\sb 0)$
are prohibited because $\langle\alpha,\check\zeta\ast\beta(\cdot,t)\rangle$
are bounded for any $\alpha\in C\sp\infty\sb{0}(\R^n)$,
while the inclusion $b(x)\in H\sp 1(\R)$
is due to $\tilde\beta\in\mathscr{S}'(\R,H\sp 1(\R))$.

Multiplying \eqref{kg-mf-beta} by $\zeta(\omega)$
and taking into account \eqref{g-at-u}, \eqref{b-b},
and the relation $\omega\sb 0^2=m^2$,
we see that the distribution $b(x)$ satisfies the equation
\begin{equation}\label{b-r}
0=\Delta b(x)+M\rho(x).
\end{equation}
Therefore, $b(x)\not\equiv 0$
due to $M\ne 0$ and $\rho(x)\not\equiv 0$.
Coupling \eqref{b-b} with $\rho$ and using \eqref{b-r}, we get:
\begin{equation}
\zeta(\omega)\langle\rho,\tilde\beta(\cdot,\omega)\rangle
=\delta(\omega-\omega\sb 0)
\langle\rho,b\rangle
=-\delta(\omega-\omega\sb 0)
\frac{\langle\Delta b,b\rangle}{M}\ne 0,
\end{equation}
since $b\in H\sp 1(\R^n)$ is nonzero.
This finishes the proof of Lemma~\ref{lemma-m}.
\end{proof}

Lemmas~\ref{lemma-same-supp} and \ref{lemma-m}
allow us to conclude that
$
\supp\tilde g(\omega)\subset
\supp\langle\rho,\tilde\beta(\cdot,\omega)\rangle,
$
finishing the proof of Proposition~\ref{prop-b-g}.
\end{proof}

\subsection{The Titchmarsh argument}

Finally, we reduce the spectrum of $\gamma(t)$ to one point
using the spectral inclusion from Proposition~\ref{prop-b-g}
and the Titchmarsh Convolution Theorem.

\begin{lemma}\label{lemma-b-point}
$\langle\rho,\beta(\cdot,t)\rangle=0$
or
$\supp\langle\rho,\tilde\beta(\cdot,\omega)\rangle=\{\omega\sb{+}\}$,
for some $\omega\sb{+}\in[-m,m]\cup Z\sb\rho$.
\end{lemma}

\begin{proof}
Denote
\begin{equation}
\gamma(t)=\langle\rho,\beta(\cdot,t)\rangle.
\end{equation}
By \eqref{f-is-such},
$
g(t):=F(\gamma(t))
=-\sum\sb{n=1}\sp{p}2 n u\sb{n}\abs{\gamma(t)}\sp{2n-2}\gamma(t).
$
Then, by the Titchmarsh Convolution Theorem,
\begin{equation}\label{spec-g-sup-00}
\sup\,\supp\tilde g
=
\max\sb{n\in\{n\le p,\,u\sb n\ne 0\}}
\sup\,\supp
\underbrace{(\tilde{\bar \gamma}\ast\tilde \gamma)
\ast\dots\ast(\tilde{\bar \gamma}\ast\tilde \gamma)}\sb{n-1}
\ast\tilde \gamma
=
p\sup\,\supp\tilde \gamma+(p-1)\sup\,\supp\tilde{\bar \gamma}.
\end{equation}

\begin{remark}
The Titchmarsh Convolution Theorem applies
because
$\supp\tilde\gamma\subset[-m,m]\cup Z\sb\rho$,
and hence is compact.
\end{remark}

Noting that
$\sup\supp\tilde{\bar \gamma}
=-\inf\supp\tilde{\gamma}$,
we rewrite \eqref{spec-g-sup-00} as
\begin{equation}\label{spec-g-sup-0}
\sup\,\supp\tilde g
=\sup\tilde \gamma+(p-1)(\sup\supp\tilde \gamma-\inf\supp\tilde \gamma).
\end{equation}
Taking into account Proposition~\ref{prop-b-g}
and \eqref{spec-g-sup-0},
we get the following relation:
\begin{equation}\label{spec-g-sup}
\sup\supp\tilde \gamma
\ge
\sup\supp\tilde g
=\sup\supp\tilde \gamma
+(p-1)(\sup\supp\tilde \gamma-\inf\supp\tilde \gamma).
\end{equation}
This is only possible if $\supp\tilde \gamma\subset\{\omega\sb{+}\}$,
for some $\omega\sb{+}\in[-m,m]\cup Z\sb\rho$.
\end{proof}

\subsection{Conclusion of the proof of Theorem~\ref{main-theorem}}

We need to prove \eqref{claim-beta}.
As follows from Lemma~\ref{lemma-b-point},
$\tilde \gamma(\omega)$ is a finite linear combination
of $\delta(\omega-\omega\sb{+})$ and its derivatives.
As the matter of fact, the derivatives could not be present
because of the boundedness of
$\gamma(t):=\langle\rho,\beta(\cdot,t)\rangle$
that follows from
\eqref{beta-beta}.
Therefore,
$\tilde \gamma=2\pi C\,\delta(\omega-\omega\sb{+})$,
with some $C\in\C$.
This implies the following identity:
\begin{equation}\label{eide}
\gamma(t)=Ce^{-i\omega\sb{+}t},
\qquad
C\in\C,\quad
t\in\R.
\end{equation}
It follows that $\tilde g(\omega)=2\pi C\,\delta(\omega-\omega\sb{+})$,
$C\in\C$,
and the representation
\eqref{ber} implies that
$\beta(x,t)=\beta(x,0)e\sp{-i\omega\sb{+} t}$.
Due to equation \eqref{kg-mf-beta} and the bound \eqref{beta-beta},
$\beta(x,t)$ is a solitary wave solution.
This completes the proof of Theorem~\ref{main-theorem}.

\section{Multifrequency solutions}
\label{sect-counterexample}

Now we consider the situation when Assumption~\ref{ass-rho} is violated.
In this case, we show that there could exist
multifrequency solutions,
indicating that the set of all (one-frequency) solitary waves
is only a proper subset of the global attractor.

Fix $\omega\sb 1\in(m,3m)$.
Set
$\omega\sb 0=\omega\sb 1/3$
and pick $\rho\in\mathscr{S}(\R^n)$ such that
the following conditions are satisfied:
\begin{equation}\label{ce-c1}
\hat\rho\at{\abs{\xi}=\sqrt{\omega\sb 1^2-m^2}}=0,
\end{equation}
\begin{equation}\label{ce-c2}
\sigma(\omega\sb 1)
:=\frac{1}{(2\pi)^n}\int\sb{\R^n}
\frac{\abs{\hat\rho(\xi)}^2\,d\sp n\xi}{\xi^2+m^2-\omega\sb 1^2}
=0.
\end{equation}

These two equalities imply that
$\sigma(\omega)$ vanishes at a certain point of $Z\sb\rho$,
violating Assumption~\ref{ass-rho}.

\begin{lemma}
There exist $a\in\R$, $b<0$ so that
equation \eqref{kg-mf} with the nonlinearity
\[
F(z)=a z+b\abs{z}^2 z,
\qquad z\in\C,
\]
admits multifrequency solutions
$\psi\in C(\R,H\sp 1)$
of the form
\[
\psi(x,t)=\phi\sb 0(x)\sin{\omega\sb 0 t}+\phi\sb 1(x)\sin{\omega\sb 1 t},
\qquad
\quad
\omega\sb 0=\frac{\omega\sb 1}{3},
\quad\phi\sb 0,\,\phi\sb 1\in H\sp 1(\R^n),
\]
with both $\phi\sb 0$ and $\phi\sb 1$ nonzero.
\end{lemma}

\begin{proof}
To make sure that the nonlinearity does not produce
higher frequencies, we assume that
\begin{equation}\label{r-p-0}
\langle\rho,\phi\sb 1\rangle=0.
\end{equation}
Due to this assumption,
\[
F(\langle\rho,\psi\rangle)
=F(\langle\rho,\phi\sb 0\rangle\sin{\omega\sb 0 t})
=a \langle\rho,\phi\sb 0\rangle\sin{\omega\sb 0 t}
+b \langle\rho,\phi\sb 0\rangle^3
\frac{3\sin{\omega\sb 0 t}-\sin{3\omega\sb 0 t}}{4}.
\]
Collecting the terms with
the factors of $\sin{\omega\sb 0 t}$ and
$\sin{\omega\sb 1 t}=\sin{3\omega\sb 0 t}$,
we rewrite the equation
$\ddot\psi=\Delta\psi-m^2\psi+\rho F(\langle\rho,\psi\rangle)$
as two following equalities:
\begin{equation}\label{ce-omega1}
-\omega\sb 0^2\phi\sb 0=\Delta\phi\sb 0- m^2\phi\sb 0
+\rho(x)
\Big(
a \langle\rho,\phi\sb 0\rangle
+\frac{3 b \langle\rho,\phi\sb 0\rangle^3}{4}
\Big),
\end{equation}
\begin{equation}\label{ce-omega3}
-\omega\sb 1^2\phi\sb 1=\Delta\phi\sb 1-m^2\phi\sb 1
-\rho(x)
\frac{b \langle\rho,\phi\sb 0\rangle^3}{4}.
\end{equation}
We define $\phi\sb 0(x)$ by
$
\hat\phi\sb 0(\xi)
=\frac{\hat\rho(\xi)}{\xi^2+m^2-\omega\sb 0^2}.
$
Since $m^2-\omega\sb 0^2>0$,
there is the inclusion $\phi\sb 1\in H\sp 1(\R^n)$.
Moreover,
\[
\langle\rho,\phi\sb 0\rangle
=\frac{1}{(2\pi)^n}\int\sb{\R^n}
\frac{\abs{\rho(\xi)}^2\,d\sp n\xi}{\xi^2+m^2-\omega\sb 0^2}
=\sigma(\omega\sb 0)>0,
\]
due to strict positivity of $\sigma(\omega)$
for $\abs{\omega}<m$
(see \eqref{def-sigma}).
Hence, for any $b$
(we take $b<0$ to comply with Assumption~\ref{ass-f}),
we may pick $a$
such that \eqref{ce-omega1} is satisfied.
We then
use \eqref{ce-omega3}
to define the function $\phi\sb 1(x)$ by
\[
\hat\phi\sb 1(\xi)
=
-\frac{ b\langle\rho,\phi\sb 0\rangle^3}{4}
\frac{\hat\rho(\xi)}{\xi^2+m^2-\omega\sb 1^2}
=
-\frac{b\sigma(\omega\sb 0)^3}{4}
\frac{\hat\rho(\xi)}{\xi^2+m^2-\omega\sb 1^2}.
\]
Due to \eqref{ce-c1},
$\phi\sb 1\in H\sp 1(\R^n)$.
We are left to check that
$\phi\sb 0$ satisfies
the assumption \eqref{r-p-0}.
Indeed, due to \eqref{ce-c2},
\[
\langle\rho,\phi\sb 1\rangle
=
-\frac{b\sigma(\omega\sb 0)^3}{4}
\frac{1}{(2\pi)^n}\int\sb{\R^n}
\frac{\abs{\hat\rho(\xi)}^2\,d\sp n\xi}{\xi^2+m^2-\omega\sb 1^2}
=0.
\]

\end{proof}

\appendix

\section{Appendix: Global well-posedness}
\label{sect-existence}

The global existence stated in Theorem~\ref{theorem-well-posedness}
is obtained by standard arguments
from the contraction mapping principle.
To achieve this, we use the integral representation
for the solutions to the Cauchy problem \eqref{kg-mf-cp}:
\begin{equation}\label{integral-representation}
\Psi(t)
=W\sb 0(t)\Psi\sb 0
+{Z}[\Psi](t),
\qquad
{Z}[\Psi](t):=
\int\sb 0\sp{t}
W\sb 0(t-s)
{\scriptsize
\Big[
\!\!
\begin{array}{c}0
\\
\rho\,F(\langle\rho,\psi(\cdot,s)\rangle)
\end{array}
\!\!
\Big]
}
\,ds,
\qquad
\Psi={\scriptsize
\Big[
\!\!\begin{array}{c}\psi\\\pi\end{array}
\!\!
\Big]
},
\quad t\ge 0.
\end{equation}
Here $W\sb 0(t)$
is the dynamical group for the linear Klein-Gordon equation
which is a unitary operator in
the space $\E\sp{-\varepsilon}$ for any $\varepsilon\ge 0$.
The bound
\begin{equation}\label{was-lemma-bounds}
\norm{{Z}[\Psi\sb 1](t)-{Z}[\Psi\sb 2](t)}\sb{\E\sp{-\varepsilon}}
\le
C\abs{t}\sup\sb{s\in[0,t]}
\norm{\Psi\sb 1(s)-\Psi\sb 2(s)}\sb{\E\sp{-\varepsilon}},
\qquad C>0,
\quad \abs{t}\le 1,
\quad
\varepsilon\ge 0,
\end{equation}
which holds for any two functions
$\Psi\sb 1$,
$\Psi\sb 2\in C(\R,\E)$,
shows that $Z[\psi]$ is a contraction operator
in $C([0,t],\E\sp{-\varepsilon})$,
$\varepsilon\ge 0$,
if $t>0$ is sufficiently small.

The contraction mapping theorem
based on the bound \eqref{was-lemma-bounds} on the nonlinear term
allows us to prove the existence and uniqueness of
a local solution in $\E$,
as well as the continuity of the map $W(t)$
(continuity with respect to the initial data).
The continuity of $W(t)$ in $\E\sb{loc}$
follows from its continuity in $\E$ and
the finite speed of propagation.

The conservation
of the values of the energy and charge functionals,
$\mathcal{H}$ and $\mathcal{Q}$,
is obtained by approximating the initial data in $\E$
with smooth initial data and using the continuity of $W(t)$
in $\E$.
For the proof of the a priori bound \eqref{eb},
we use \eqref{bound-below}
to bound $\norm{\Psi}\sb{\E}$
in terms of the value of the Hamiltonian:
\begin{equation}\label{t-bound-1}
\norm{\Psi}\sb{\E}\sp 2
\le\frac{2m^2}{m^2-2{B}\norm{\rho}\sb{L\sp 2}^2}
\left(\mathcal{H}(\Psi)-{A}\right),
\qquad\Psi\in\E.
\end{equation}
This bound
allows us to extend the existence results for all times,
proving the global well-posedness of
\eqref{kg-mf-cp} in the energy space.

Finally,
the continuity of $W(t)$
in $\E\sp{-\varepsilon}$ and $\E\sp{-\varepsilon}\sb{loc}$,
$\varepsilon\ge 0$,
follows from
the contraction mapping theorem
(based on \eqref{was-lemma-bounds})
and the finite speed of propagation.


\def\cprime{$'$} \def\cprime{$'$}

\end{document}